\documentclass[12pt]{iopart}
\usepackage{graphicx}

\begin{document}

\title[Generation and propagation of entanglement]{Generation and propagation of  entanglement in driven coupled-qubit systems}

\author{J Li and G~S~Paraoanu}

\address{Low Temperature Laboratory, Helsinki University of Technology,
P.O. Box 5100, FIN-02015 TKK, Finland}

\ead{paraoanu@cc.hut.fi}


\begin{abstract}
In a bipartite system subject to decoherence from two separate reservoirs, the entanglement is typically destroyed faster than
for single reservoirs. Surprisingly however, the existence of separate reservoirs can also have a beneficial entangling effect: if the qubits are coupled and driven externally by a classical field, the system ends up in a stationary state characterized by a finite degree of
entanglement. This phenomenon occurs only in a certain region of the parameter space and the structure of the stationary state has a universal form which does not depend on the initial state or on the specific physical realization of  the qubits.
We show that the entanglement thus generated can be propagated within a quantum network using simple local unitary operations. We suggest the use of such systems as "batteries of entanglement" in quantum circuits.

\end{abstract}
\submitto{\NJP}

\maketitle

\section{Introduction}

When two qubits  are exposed to uncorrelated weak noise, qualitatively new phenomena, known as sudden death of entanglement, have been predicted
theoretically \cite{Yu1,Yu3} and confirmed experimentally for photons \cite{almeida} and atoms \cite{atoms}. These phenomena include the abrupt (rather than exponential) decay of concurrence for certain initial entangled states \cite{Yu1} and the non-additivity of decoherence rates \cite{Yu3}. They indicate that two different environments have a detrimental effect on the entanglement, which gets stronger with the addition of external influences, as shown by the case of driven qubits \cite{driven} and finite-temperature baths \cite{finitetemp}. This contrasts to the case of a single environment, where the constructive role of decoherence has been known for some time \cite{braun}. Other cases in which a single environment can in fact support the occurence of entanglement have been pointed out in various contexts: for example, if the qubits are allowed to exchange excitations via a third continuously-monitored quantum object \cite{third}, if the qubits are driven \cite{milburn}, dipole-coupled \cite{dipole}, or if vacuum fluctuations (rotating terms) are not negligible \cite{ng}.

In this paper we show that, for coupled and driven qubits interacting with two different reservoirs, entanglement can be generated and maintained at long time-scales.  The existence of this effect does not depend on the specific physical realization of the qubits but requires the presence of all three ingredients: dissipation, driving, and coupling. We show that steady-state generation of entanglement is possible in a region which is outside the range of validity of the secular approximation \cite{book}, where the concurrence can reach a maximal universal value (half the inverse of the golden ratio) if a certain simple relation between the driving field and the coupling is established. In a wider quantum-information context, this effect  could find applications such as on-demand creation of complex entangled states by manipulating the dissipation \cite{zoller}, a process which  could be regarded as a form of adiabatic quantum computing \cite{cirac}. Here  we suggest that, since stationary entanglement is robust under decoherence, independent on the initial state of the qubits, relatively easy to generate, and has a system-independent value, such systems could be used as "entanglement batteries", and we show that, using simple local unitary operations available now for systems such as superconducting qubits,  this entanglement can be harvested and transmitted further in a quantum circuit.

\section{Generation of entanglement}

We consider two qubits of Larmor frequency $\nu_j$ ($j=1,2$), coupled by a dipole-dipole interaction of strength $\omega^{xx}(t)\ll\nu_j$, and irradiated by external monochromatic fields of frequency $\omega_{j}$ and vacuum Rabi frequencies $\Omega_{j}$. The two states of each qubit are denoted by $|0\rangle$ and $|1\rangle$, and for the Bell basis of the two qubits we will use the notation $|\Psi^{\pm}\rangle = 1/\sqrt{2}(|01\rangle\pm|10\rangle)$, and  $|\Phi^{\pm}\rangle = 1/\sqrt{2}(|00\rangle\pm|11\rangle)$. The dipole-dipole coupling between the qubits can be either direct or resulting from virtual excitations through a third object, as it is the case in many qubit architectures \cite{ind}.  In the Schr\"odinger picture and with $\hbar =1$,

\begin{eqnarray}
&& H^{(S)} = \sum_{j=1,2} \frac{\nu_j}{2}\sigma_j^z + \omega^{xx}(t)\sigma_1^x\sigma_2^x + \sum_{j = 1,2} \Omega_j\cos(\omega_j t)\sigma_j^{x},
\label{eq_hamiltonians}
\end{eqnarray}
where, to account for the situation in which the qubits have very different Larmor frequencies \cite{bertet}, we have considered a modulated coupling
$\omega^{xx} (t)= 2 \omega^{xx} \cos [(\omega_{2} -\omega_{1})t]$.
The dissipation is described by the standard Born-Markov master equation \cite{book},
\begin{eqnarray}
\dot{\rho^{(S)}} = -i [H^{(S)}, \rho^{(S)}] + {\cal L}[\rho^{(S)}] , \label{eq_original_master_equation}
\end{eqnarray}
where the  Liouvillean
\begin{eqnarray}
{\cal L}[\rho^{(S)}] = \sum_{j=1,2} \frac{\Gamma_j}{2} \left( 2\sigma_j^-
\rho^{(S)} \sigma_j^+ - \sigma_j^+ \sigma_j^-
\rho^{(S)} - \rho^{(S)} \sigma_j^+
\sigma_j^- \right)
\end{eqnarray}
models the longitudinal dampings of the qubits, and $\Gamma_j$ is the standard energy relaxation rate.

We  work in a rotating reference frame, characterized by the transformation
$R = \exp[i(\omega_1\sigma_1^z + \omega_2\sigma_2^z)t / 2]$. Using the identities $R^\dag \sigma_{j}^{\pm}R = \sigma_{j}^{\pm}
\exp (\mp i\omega_{j}t)$ and eliminating the fast counter-rotating terms {\it via} the rotating wave approximation, we obtain a time-independent master equation in the rotating frame,

\begin{equation}
\dot{\rho} = -i [H, \rho] + {\cal L}[\rho] , \label{eq_master_equation_in_rf}
\end{equation}
with $\rho = R\rho^{(S)} R^\dag$, and
\begin{equation}
H \approx \sum_{j=1,2}\left( \frac{\delta_j}{2}\sigma_j^z + \frac{\Omega_j}{2}\sigma_j^x \right) +
\frac{\omega^{xx}}{2}\left( \sigma_1^x\sigma_2^x + \sigma_1^y\sigma_2^y\right). \label{eq_h_rf}
\end{equation}
Here $\delta_j = \nu_j - \omega_j$ are the detunings of the qubits from the corresponding driving frequencies. In order to use the rotating wave approximation, we have taken $\omega_{1}, \omega_{2}, |\omega_{1}-\omega_{2}|\gg \Omega_{j}, \delta_{j}, \omega^{xx}, \Gamma_{j}$. For the case of qubits with close Larmor frequencies, there is no need  to modulate the coupling: an effective Hamiltonian of the type Eq. (\ref{eq_h_rf}) can be obtained \cite{driven,arx} by the same transformations and working at equal driving frequencies $\omega_{1}=\omega_{2}$.
From now on, we will  also refer to the interaction part (containing $\omega^{xx}$) of the Hamiltonian Eq. (\ref{eq_h_rf}) as $H^{xx}$, and the rest as
$H_0$ ($H = H_{0} + H^{xx}$).

\begin{figure}[htb]
\includegraphics[width=9cm]{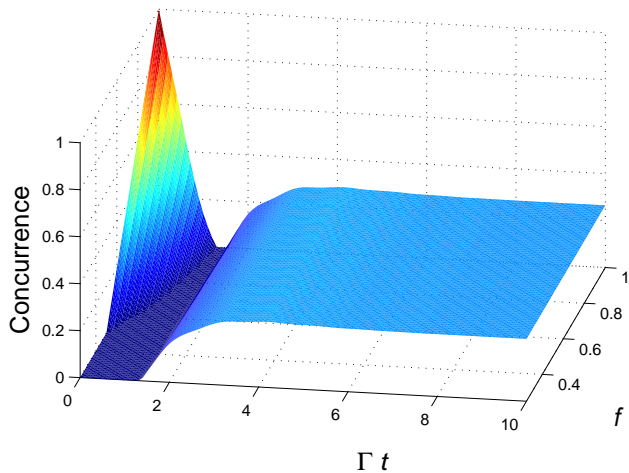}
\caption{(color online). Time evolution of the concurrence for a Werner state, with the system parameters $\Omega = 2\Gamma$ and $\omega^{xx} = 5\Gamma$.}
\label{fig_recreation}
\end{figure}

To explore the entanglement properties of this system, we have first solved numerically Eq. (\ref{eq_master_equation_in_rf}) for $\delta_j =\delta$, $\Omega_j = \Omega$, and $\Gamma_{j}=\Gamma$. The entanglement between the qubits at any time is characterized by Wooters' concurrence \cite{wootters}, defined as
${\rm max}\{ 0, {\cal C}\}$, where ${\cal C}=\lambda_1-\lambda_2-\lambda_3-\lambda_4$,
 and $\lambda_i$s are the eigenvalues of $\sqrt{\sqrt{\rho} \widetilde{\rho}\sqrt{\rho}}$
in decreasing order, with $\widetilde{\rho}\equiv (\sigma^y \otimes \sigma^y) \rho^\ast (\sigma^y \otimes \sigma^y)$. We find that for most of the values of $\Omega$ and $\omega^{xx}$ either sudden death of entanglement or exponential decay of entanglement occurs, depending on the initial state \cite{Yu1}. However, for certain values of the driving field $\Omega$ and  coupling $\omega^{xx}$ a process of concurrence buildup occurs, as shown in Fig. \ref{fig_recreation} for the case of initial Werner states \cite{brassard}, defined as $\rho_{W} = [(1-f)/3] I + [(4f-1)/3]|\Psi^{-}\rangle \langle\Psi^{-}|$.

\begin{figure}[htb]
\includegraphics[width=5cm]{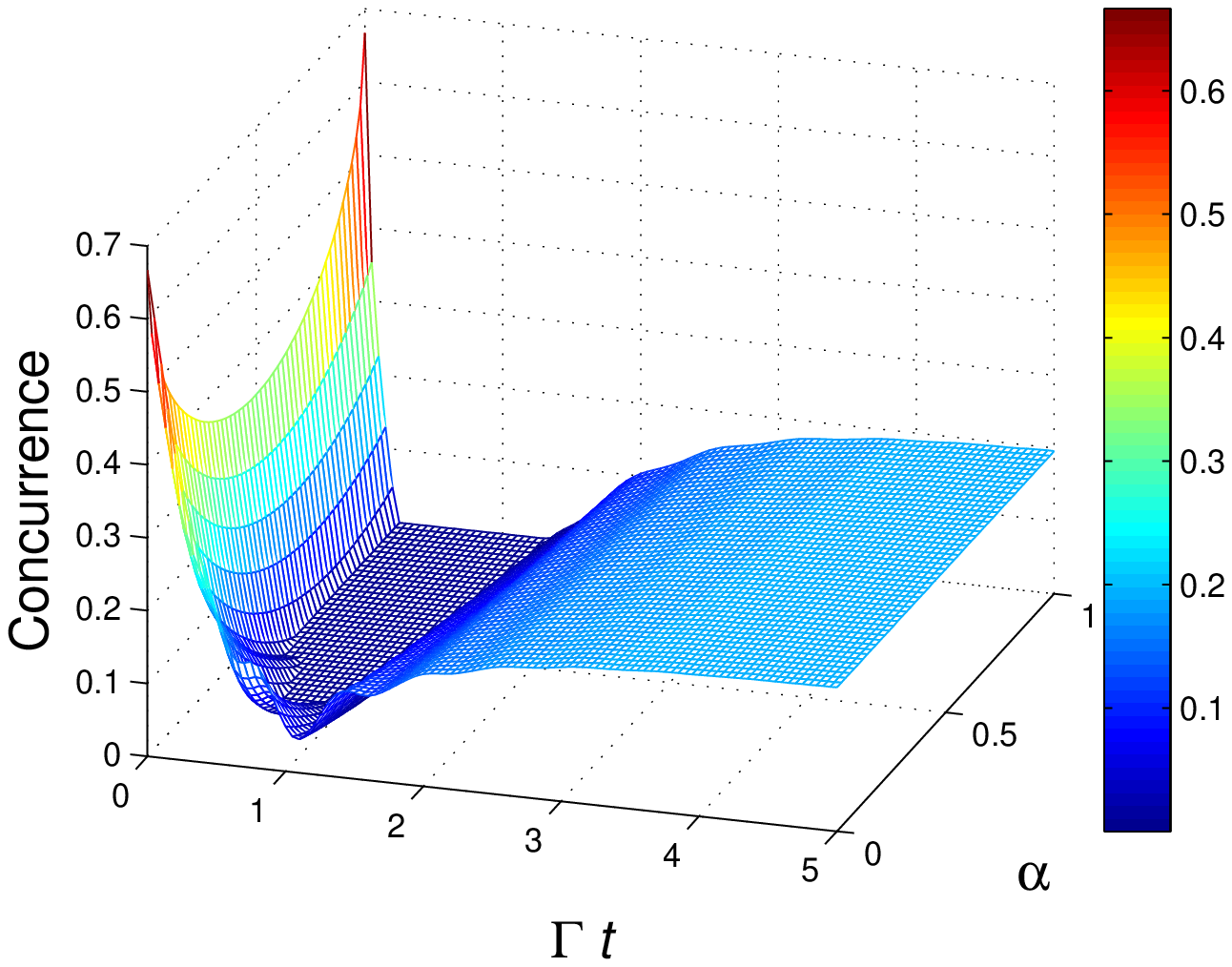}
\includegraphics[width=5cm]{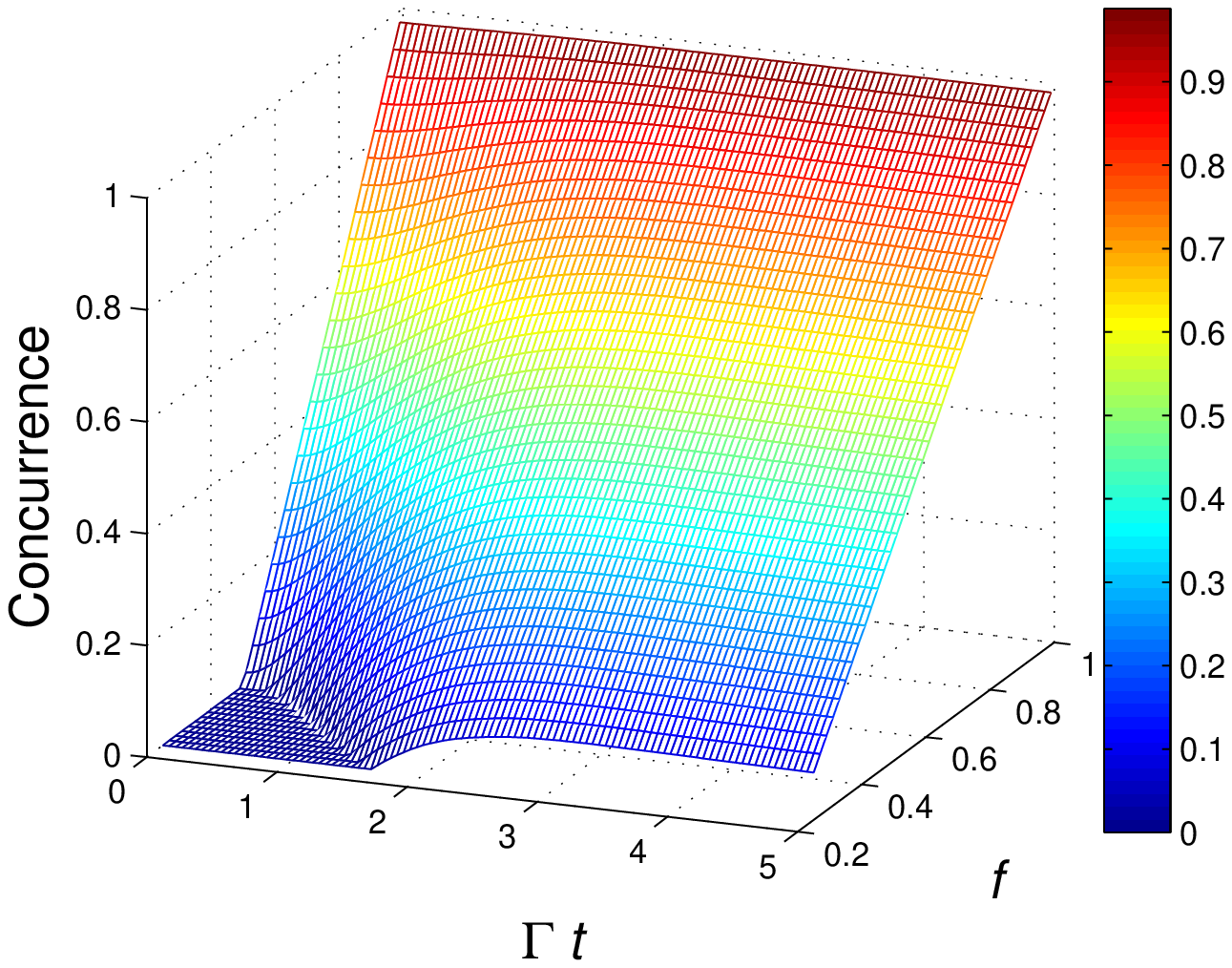}
\includegraphics[width=5cm]{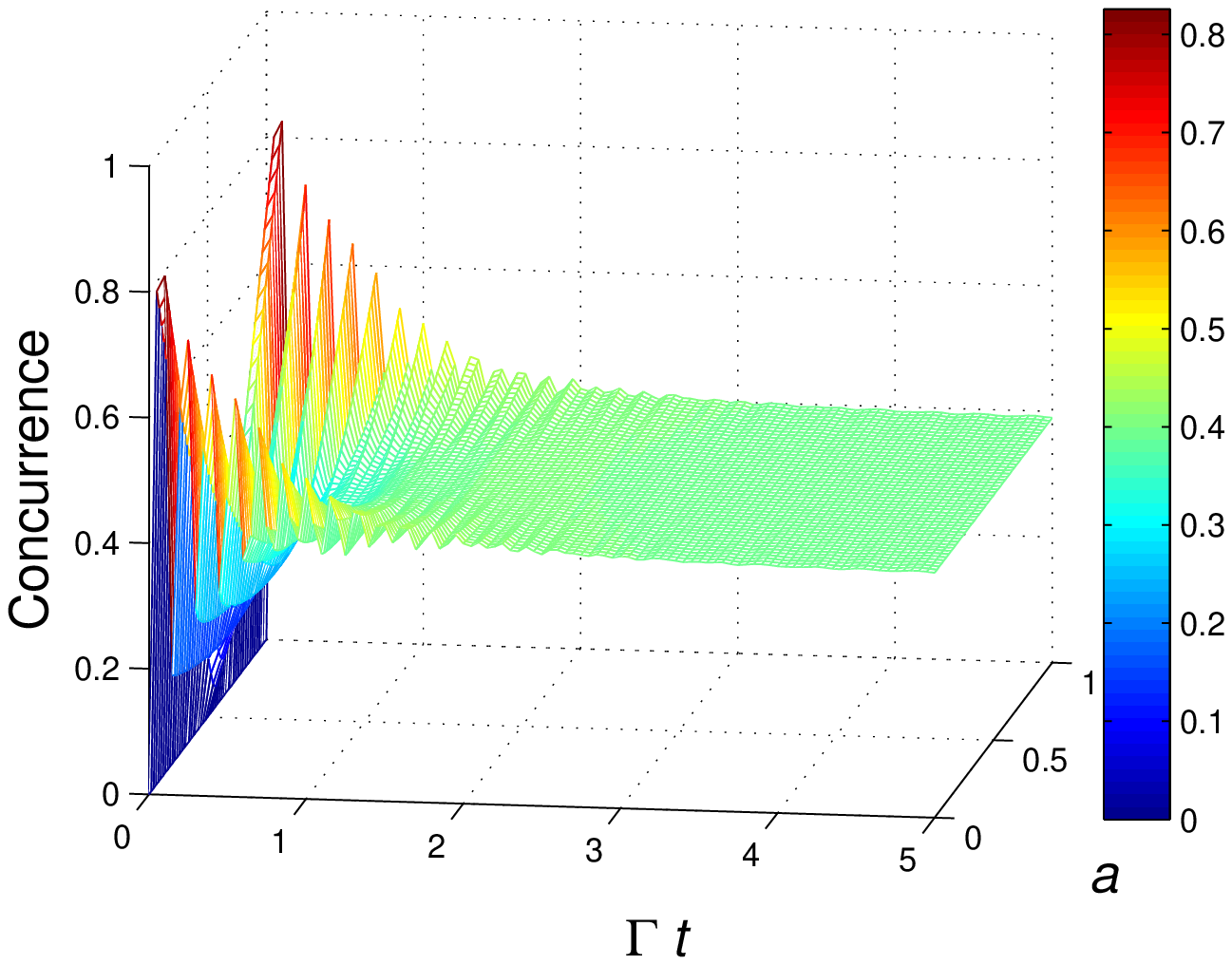}
\caption{(color online). Concurrence evolution for (left) YE states, for (middle) Werner states, and (right) for $eg-ge$ states. For all the plots, $\omega^{xx}=10\Gamma$, $\Omega=1.5\Gamma$.}
\label{common}
\end{figure}

We first note that this effect is  qualitatively different
from the case of a single reservoir, with Liouvillean superoperator
\begin{eqnarray}
{\cal L}_{single}[\rho] =
\frac{\Gamma}{2} \left( 2 S^{-} \rho S^{+} - S^{+}S^{-}
\rho - \rho S^{+}S^{-}\right),
\end{eqnarray}
with $S^{\pm}=\sum_{j=1,2} \sigma_j^{\pm}$  being collective spin operators.
In this situation, the emergence of an entangled steady state {\it via}
evolution
depends on the initial state, as shown in Fig. \ref{common}, where we plotted the time-dependent concurrence for three classes of states, Werner states, Yu-Eberly (YE) states \cite{Yu2}
(states of the type $\rho_{YE}= (2/3)|\Psi^{+}\rangle\langle\Psi^{+}| + [(1-\alpha)/3]|11\rangle\langle 11| + (\alpha /3 ) |00\rangle \rangle 00 |$)
, and states diagonal in the subspace with one excitation, which we denote by
$\rho_{eg-ge}=(1-a)|10\rangle\langle 10| + a|01\rangle\langle 01|$. The reason for this is the existence, for common reservoirs, of a decoherence-free subspace: as a result, the amount of entanglement contained in the  asymptotic state  depends in general on the projection of the initial state onto this subspace \cite{an}. A number of effects related to decoherence-free
subspaces in interacting qubits have been studied recently: for example, it has been shown how to use the qubit-qubit coupling to effectively produce a low-decoherence subspace even when the qubits have additional separate decoherence channels \cite{you}, and how to create maximally-entangled state by suitable measurements on the decay photons \cite{third}.

\begin{figure}[htb]
\includegraphics[width=10cm]{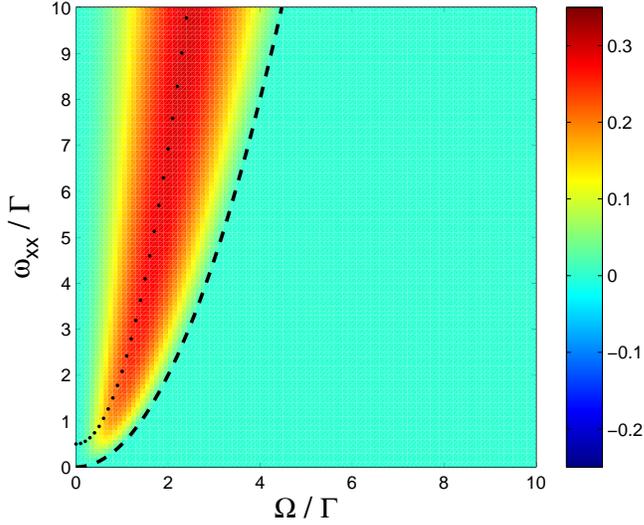}
\caption{(color online). Numerical values for the concurrence at a time $t=10^{3}\Gamma^{-1}$, as a function of $\Omega$ and $\omega^{xx}$. The dashed line is a plot of the parabola Eq. (\ref{zzero}) and the dotted line is a plot of Eq (\ref{max}). }
\label{fig_phase_diagram}
\end{figure}

We now return to the case of two reservoirs. Entanglement generation occurs only in a certain region of the parameter space: in
Fig. \ref{fig_phase_diagram} we have plotted the concurrence for a long time-scale
evolution  $t\gg \Omega^{-1}, \Gamma^{-1}$. The maximum value of the concurrence is reached at resonance ($\delta =0$) and it increases slowly with $\Omega /\Gamma$ and $\omega^{xx}/\Gamma$. Although intuitively  one might expect that increasing either the coupling or the pumping strength would increase the steady-state concurrence, this is not the case: both the ratios $\Omega /\Gamma$ and $\omega^{xx} /\Gamma$ are important, showing that all three processes (coupling, pumping, and decoherence) contribute to this effect. It also means that the  secular approximation \cite{book}
is not valid in the region of the parameter space where the effect occurs.

To get some insight into the mechanism responsible for the generation of steady-state concurrence, we will take a perturbative approach: if there is no coupling between the two qubits, then in the stationary regime the bipartite system is described by the separable density matrix $\rho^{(\infty)}_{1}\rho^{(\infty)}_{2}$. Consider a single qubit $j$ under constant on-resonance driving: in the rotating frame,  the master equation is simply
\begin{eqnarray}
\dot{\rho_{j}} = -\frac{i\Omega}{2}[\sigma^x,\rho_{j}]  + \frac{\Gamma}{2}\left( 2\sigma^-\rho_{j}\sigma^+ - \sigma^+\sigma^-\rho_{j} - \rho_{j}\sigma^+\sigma^- \right) , \label{eqnnb}\ \
\end{eqnarray}
and the steady-state is (see \ref{appendixa})
\begin{equation}
\rho_{j}^{(\infty)} = \frac{1}{\Gamma^2 + 2\Omega^2}
\left(\begin{array}{cc} \Omega^2 & i\Omega\Gamma \\ -i\Omega\Gamma & \Gamma^2 + \Omega^2 \end{array}\right).\label{hoi}
\end{equation}

We now aim at finding the first-order contribution
in $\omega^{xx}/\Gamma \ll 1$ to the density matrix and the concurrence.
We write the solution as  $\rho^{(\infty )}\approx\rho_{1}^{(\infty)}\rho_{2}^{(\infty)}+\rho^{xx}$, where, in order to ensure a unit value for $\rho^{(\infty )}$, we search for a traceless density matrix
$\rho^{xx}$  which satisfies the equation $-i[H_{0}, \rho^{xx}] -i [H^{xx}, \rho_{1}^{(\infty )}\rho_{2}^{(\infty )}] + {\cal L}{\rho^{xx}} = 0$. Here we neglect
the second-order contribution in $\omega^{xx} /\Gamma$ coming from the commutator $[H^{xx}, \rho^{xx}]$. With the notation $\rho^{xx}_{jk,lm}=\langle jk|\rho^{xx}|lm\rangle$, we find that $\rho^{xx}$ has matrix elements
$\rho^{xx}_{11,00} = \rho^{xx*}_{00,11}= 2i\Omega^{2}\Gamma \omega^{xx} /(\Gamma^2 + 2 \Omega^2 )^2$, $\rho^{xx}_{10,00} = \rho^{xx*}_{00,10}= \rho^{xx}_{01,00}= \rho^{xx*}_{00,01}= 2\Omega\Gamma^{2} \omega^{xx} /(\Gamma^2 + 2 \Omega^2 )^2$ and the rest zero. The quantity ${\cal C}$ associated to the density matrix $\rho_{1}^{(\infty)}\rho_{2}^{(\infty)}+\rho^{xx}$ is
${\cal C}=2\Omega^{2} (2\omega^{xx}\Gamma-\Omega^{2})(\Gamma^2 + 2\Omega^2 )^{-2}$,
and we have checked that it gives excellent fits to the numerical results in the region where $\omega^{xx}/\Gamma \ll 1$ for both negative and positive values of ${\cal C}$; it also predicts that ${\cal C}$ will become positive
for $\omega^{xx} = \Omega^{2} /2\Gamma$, in  agreement with the exact result derived below Eq. (\ref{zzero}).

Although in the region of interest (finite, large concurrence) the problem is clearly nonperturbative,
the perturbative solution serves as a heuristic guide to find the stationary  density matrix that solves Eq. (\ref{eq_master_equation_in_rf}) under the condition $\dot{\rho} = 0$. Note that this condition implies solving for the 15 unkown real entries of $\dot{\rho} = 0$. Surprisingly, although the algebraic calculations are rather complicated, at resonance we find a simple solution for $\rho^{(\infty )}$,
\begin{eqnarray}
\rho^{(\infty)} &=& \frac{(\Gamma^2 + 2\Omega^2 )^2}{[(\Gamma^2 + 2\Omega^2)^2  + 4\omega^{xx2}\Gamma^2 ]^2}
\left(\rho_{1}^{(\infty)}\rho_{2}^{(\infty)} + \rho^{xx} \right) \nonumber \\
&& +
\frac{\Gamma^2\omega^{xx2}}{[(\Gamma^2 + 2\Omega^2)^2  + 4\omega^{xx2}\Gamma^2 ]^2}(1-\sigma^{(1)}_{z})(1-\sigma^{(2)}_{z}). \nonumber
\end{eqnarray}
This solution preserves the structure of the perturbative case - note that the role of the second  term is just to ensure that $Tr\rho^{(\infty)} =1$, due to the appearance of the prefactor in front of $\rho_{1}^{(\infty)}\rho_{2}^{(\infty)} + \rho^{xx}$ in the first term.
Similarly, ${\cal C}$ can be calculated analytically and the result resembles the nonperturbative case,
\begin{equation}
{\cal C}= 2\Omega^{2}\frac{2\omega^{xx}\Gamma -\Omega^2}
{(\Gamma^2 + 2\Omega^2)^2  + 4\omega^{xx2}\Gamma^2 }.
\end{equation}

The crossover between the region of finite steady-state concurrence and the region of zero concurrence is given by the parabola
\begin{equation}
\omega_{0}^{xx} = \frac{\Omega^2}{2\Gamma},\label{zzero}
\end{equation}
and the concurrence reaches a maximum value of
\begin{equation}
{\cal C}_{\rm max} = \frac{\Omega^2}{\Omega^{2}+\sqrt{\Omega^{4}+(\Gamma^{2} + 2\Omega^{2})^2}} ,
\end{equation}
for
\begin{equation}
\omega_{\rm max}^{xx}= \frac{\Omega^{2}}{2\Gamma}+\frac{1}{2\Gamma}\sqrt{(2\Omega^{2}+\Gamma^{2})^{2}+\Omega^{4}}. \label{max}
\end{equation}
At large values of the driving field $\Omega /\Gamma \gg 1$, we obtain a maximum concurrence of $1/(1+\sqrt{5})=0.309$ (which incidentally  is half of the inverse of the golden ratio) , a result which we have confirmed numerically; the density matrix corresponding to this concurrence is
\begin{equation}
\rho_{\rm max}^{(\infty )} = \frac{1}{2\sqrt{5}(2\sqrt{5}+1)} I +
\frac{1}{2\sqrt{5}} \left( \begin{array}{cccc}0& 0& 0& i \\ 0& 0& 0&0 \\ 0&0&0&0 \\-i &0&0& 1+\sqrt{5} \end{array}\right). \label{form}
\end{equation}

\begin{figure}[htb]
\includegraphics[width=9cm]{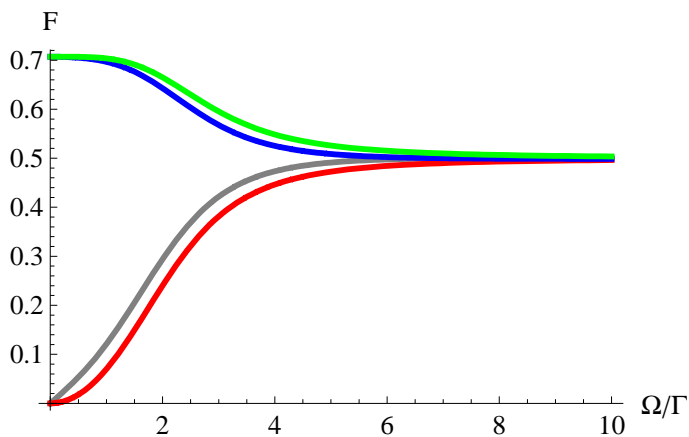}
\caption{The fidelities $F(|\psi^{+}\rangle , \rho^{(\infty )})$ (gray),
$F(|\psi^{-}\rangle , \rho^{(\infty )} ) $ (red),
$F(|\phi^{+}\rangle , \rho^{ (\infty )} )$ (blue), and
$F(|\phi^{-}\rangle , \rho^{(\infty )} )$ (green),
for $\omega^{xx} = 7\Gamma$ and $\delta = 0$.}
\label{fid}
\end{figure}

In the case of finite detuning, the calculations are more involved but it is still
possible to obtain analytical expressions. We find for ${\cal C}$
\begin{equation}
{\cal C}= \frac{2\Omega^{2}}{{\cal N}^{4}}\left(2\omega^{xx}|\tilde{\Gamma} |-\Omega^2\right),
 \label{conci}
\end{equation}
where
\begin{equation}
{\cal N}=[(|\tilde{\Gamma}|^2
+2\Omega^2 )^2 + 4\omega^{xx}|\tilde{\Gamma}|^{2} (\omega^{xx}+2\delta)| ]^{1/4},
\end{equation}

and $\tilde{\Gamma} = \Gamma + 2i\delta$. The  fidelities with respect to the Bell basis also have simple forms
$ F(|\psi^{+}\rangle , \rho^{(\infty )} ) = \Omega {\cal N}^{-2}\sqrt{ 2|\tilde{\Gamma}|^2 + \Omega^2}$,
$F(|\psi^{-}\rangle , \rho^{(\infty ) }) = \Omega^{2}{\cal N}^{-2}$, $F(|\phi^{+}\rangle , \rho^{(\infty )}) = \sqrt{1/2-\Omega^2 {\cal N}^{-4}(2\Gamma^2 + \Omega^2 - 4 \delta \omega^{xx})}$, and
$F(|\phi^{-}\rangle , \rho^{(\infty )} ) =  \sqrt{1/2-\Omega^2 {\cal N}^{-4}(2\Gamma^2 + \Omega^2 + 4 \delta \omega^{xx})}$. In Fig. \ref{fid} we plot the fidelities at resonance for various $\Omega$'s.

The analysis above shows that the mechanism of generating entanglement is related to the existence of off-diagonal matrix elements (single-qubit coherences) in $\rho_{j}^{(\infty )}$: these get coupled by the interaction (as shown by the existence of linear terms in $\Gamma$ and $\Omega$ in $\rho^{xx}$).
It is interesting to note that the role of driving is solely to pump energy in the system: the source can be a classical one and perfect coherence is not a strong requirement either. Indeed, if the qubits are driven by incoherent sources of components $k$,  $\sum_{k}\Omega^{(k)}\cos (\omega_{j} t + \varphi^{(k)})(\sigma_{1}^{x}+\sigma_{2}^{x})$, then the same formula for the concurrence Eq. (\ref{conci}) can be obtained, with
$\Omega = |\sum_{k}\Omega^{(k)}\exp (i\varphi^{(k)})|$, and, as long as this quantity is not exactly zero, the effect described above will be obtained.

\section{Propagation of entanglement in quantum circuits}

\begin{figure}[htb]
\includegraphics[width=4.5cm]{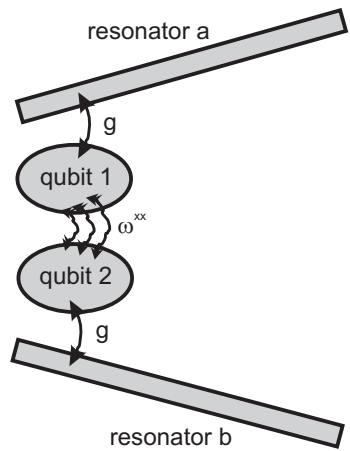}
\caption{Transmission of entanglement: schematic of the two coupled qubits and the cavities.}
\label{schematic}
\end{figure}

We now show that designing quantum circuits that would harvest and propagate this stationary entanglement at places
where it is needed in a quantum processor is possible using a tunable Jaynes-Cummings interaction, a technology already available for example in the field of superconducting qubits \cite{qubits,resonator}. Such schemes are of general interest for entanglement distribution between the nodes of a quantum networks \cite{kraus}. Since entanglement can be regarded as a resource for processing information,
such a circuit could effectively work as an "entanglement battery": when entangled fields are needed in other parts of the processor, it would be enough to outcouple them
using for example two transmission lines, similar to standard voltage or current sources, as shown schematically in Fig. \ref{schematic}.

Based on the already-achieved qubit manipulation protocols in these quantum computing architectures, we consider a circuit design in which the qubits are  coupled to each other and also to two sections of two coplanar waveguides (or striplines) used as resonating cavities. Then, after the system reaches the stationary state, and for a short time $\tau \ll \omega^{xx-1}, \Gamma^{-1}, \Omega^{-1}$ , we couple simultaneously the qubits 1,2
to the modes $a$ and $b$, respectively,  of the two cavities. Alternatively, a tunable interqubit coupling can be used to turn off $\omega^{xx}$ \cite{science}. The coupling strength (vacuum Rabi frequency) $g$
of each qubit-resonator system is chosen such that $g\gg \omega^{xx},\Gamma, \Omega$. Then the effective
Hamiltonian describing the oscillations between a qubit and its corresponding cavity is, in the rotating frame,
of the Jaynes-Cummings type ($g(a^{+}\sigma_{1}^{-}+a^{-}\sigma_{1}^{+})$ for the first qubit, and $g(b^{+}\sigma_{2}^{-}+
b^{-}\sigma_{2}^{+})$ for the second one).

The dynamic evolution is, under the conditions specified above, such that vacuum Rabi oscillation between the qubits and the respective resonators occur, which can be described in a simple way as follows:
\begin{eqnarray}
|1_{1},0_{a}\rangle &\rightarrow &\cos (g\tau) |1_{1},0_{a}\rangle -i \sin (g\tau) |0_{1},1_{a}\rangle , \label{t1}\\
|0_{1},0_{a}\rangle &\rightarrow &|0_{1},0_{a}\rangle \label{t2}, \\
|1_{2},0_{b}\rangle &\rightarrow &\cos (g\tau) |1_{2},0_{b}\rangle -i \sin (g\tau) |0_{2},1_{b}\rangle ,\label{t3}\\
|0_{2},0_{b}\rangle &\rightarrow &|0_{2},0_{b}\rangle ,\label{t4}
\end{eqnarray}
where for clarity we have introduce the subscripts 1 and 2 to denote the two qubits.
We see that for $g\tau =\pi /2$ this transformation realizes two independent
$\pi /2$ rotations in the two-dimensional subspaces spanned by $|0_{1},1_{a}\rangle ,|1_{1},0_{a}\rangle$ and
$|0_{2},1_{b}\rangle ,|1_{2},0_{b}\rangle$ respectively.
 As a result, a maximally entangled state for example $|\Psi^{+}\rangle$
 will be transferred, after a time
$\tau$, into a photonic Bell state
\begin{equation}
|\Psi^{+}\rangle\otimes|0_{a}0_{b}\rangle \rightarrow -i|0_{1}0_{2}\rangle \otimes
\frac{1}{\sqrt{2}}(|1_{a}0_{b}\rangle + |0_{a}1_{b}\rangle ),
\end{equation}
where $|1_{a}\rangle$, and $|0_{a}\rangle$ are the two states of the cavity $a$, and similar for $b$. In general, for an arbitrary $\tau$,
the density matrix in the photon basis
$\{ |1_{a},1_{b}\rangle , |1_{a},0_{b}\rangle ,|0_{a},1_{b}\rangle , |0_{a},0_{b}\rangle \}$
can be obtained by applying the transformation Eq. (\ref{t1} - \ref{t4}) and tracing out the qubits' degrees of freedom.
The result is
\begin{equation}
\left(\begin{array}{cccc} 0 & 0 & 0 & 0  \\
0 & (1/2)\sin^2 g\tau & (1/2)\cos^2 g\tau & 0 \\
0 & (1/2) \sin^2 g\tau & (1/2)\cos^2 g\tau & 0 \\
0 & 0 &  0 & \cos^2 g\tau
\end{array}\right),
\end{equation}
with concurrence ${\cal C}$  given by  ${\cal C} = \sin^2 g\tau$.

We now apply the same procedure for the general case in which the initial density matrix is the steady-state solution
for two interacting qubits and two independent reservoirs.
If we denote the elements of this matrix by $\rho_{jk,lm}^{(\rm transmitted )}=\langle j_{a}k_{b}|\hat{\rho}^{(\rm transmitted)}
|l_{a},m_{b}\rangle$, where $j,k,l,m \in \{0,1\}$, we find, after a qubit-resonator coupling time $\tau$ and after tracing out the qubits, the following
density matrix elements (we give only the values of the 9 independent elements, the rest can be found from $Tr(\rho^{\rm (transmitted)}) = 1$ and  $\rho_{jk,lm}^{(\rm transmitted )} =  -\rho_{lm,jk}^{(\rm transmitted )*}$)
in the photon basis  $\{ |1_{a},1_{b}\rangle , |1_{a},0_{b}\rangle ,|0_{a},1_{b}\rangle , |0_{a},0_{b}\rangle \}$,
\begin{eqnarray}
 \rho_{11,11}^{\rm (transmitted)} &=& (\sin g\tau)^4 \rho_{11,11}^{(\infty )}\\
\rho_{11,10}^{\rm (transmitted)} &=& -i(\sin g\tau)^3  \rho_{11,10}^{(\infty )}\\
\rho_{11,01}^{\rm (transmitted)} &=& -i (\sin g\tau)^3 \rho_{11, 01}^{(\infty )} \\
\rho_{11,00}^{\rm (transmitted)} &=& -(\sin g\tau)^2 \rho_{11,00}^{(\infty )}  \\
\rho_{10,10}^{\rm (transmitted)} &=& (\sin g\tau)^2 \rho_{10,10}^{(\infty )} + (\sin g\tau)^2 (\cos g\tau)^2 \rho_{11,11}^{(\infty )} \\
\rho_{10,01}^{\rm (transmitted )} &=& (\sin g\tau)^2 \rho_{10,01}^{(\infty )}\\
\rho_{10,00}^{\rm (transmitted)} &=& -i (\sin g\tau) \rho_{10,00}^{(\infty )} -
i (\sin g\tau)(\cos g\tau)^2  \rho_{11,01}^{(\infty )} \\
\rho_{01,01}^{\rm (transmitted)} &=& (\sin g\tau)^2 \rho_{01,01}^{(\infty )}+ (\sin g\tau )^2 (\cos g\tau)^2\rho_{11,11}^{(\infty )} \\
\rho_{01,00}^{\rm (transmitted)} &=&
-i (\sin g\tau )^2 \rho_{01,00}^{(\infty )} -i (\sin g\tau ) (\cos g\tau)^2 \rho_{11,10}^{(\infty )}
\end{eqnarray}

\begin{figure}[htb]
\includegraphics[width=9cm]{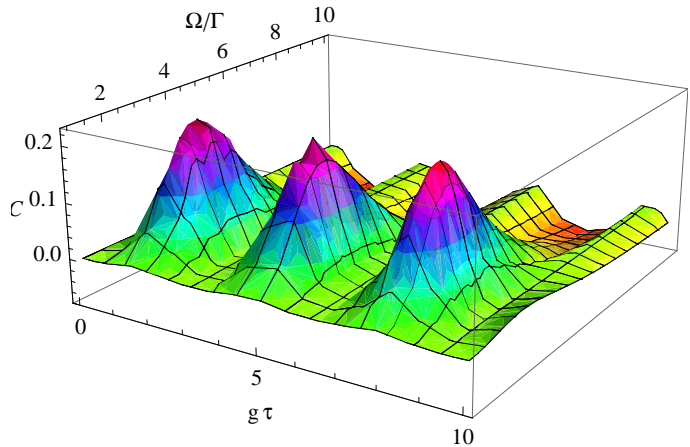}
\caption{Photon-photon concurrence as a function of coupling time $\tau$ for $\omega^{xx} = 20\Gamma$ and $\delta = 0$.}
\label{evolution}
\end{figure}

 The concurrence corresponding to this density matrix is plotted in Fig. \ref{evolution} for various qubit-resonator coupling times.
We also note that an interesting property of the procedure described above is that, if the input is a density matrix in the X-form (nonzero elements only on the  diagonal and counterdiagonal  \cite{Yu1,Yu3,almeida,driven}), the transmitted density matrix is also in the X-form.

We find, after a qubit-resonator coupling time $g\tau=\pi /2$, and after tracing out the qubits, the following
density matrix in the photon basis  $\{ |1_{a}1_{b}\rangle , |1_{a}0_{b}\rangle ,|0_{a}1_{b}\rangle , |0_{a}0_{b}\rangle \}$,
\begin{equation}
\rho^{\rm transmitted}_{jk,lm}=(-1)^{j+k}i^{j+k+l+m}\rho_{jk,lm}^{(\infty)}.
\end{equation}
This density matrix has the same concurrence as the initial one, in fact, it is identical to the initial qubit-qubit density matrix up to the local transformations $|0_{a}\rangle \rightarrow |0_{a}\rangle,
|1_{a}\rangle \rightarrow -i|1_{a}\rangle$ and similarly for photon $b$. These are phase quantum gates, and - if the exact form of the steady-state
density matrix needs to be recovered - can be implemented by a global redefinition of the phase and two separate $\pi$ rotations around the $z$-axis,
$\exp[-i\pi]\exp[-i\pi (\sigma_{a}^{z} + \sigma_{b}^{z})/2]$.

Repeating this procedure with further circuit elements, entangled pairs can be outcoupled and processed in various parts of a quantum processor \cite{nielsen}, where they can be distilled \cite{distillation}
to maximally entangled states (used for example in quantum gates) or they can be used
directly for tasks that do not require highly entangled states, such as quantum catalysis, - making possible certain local transformations and enhancing distillation algorithms \cite{plenio}. Also, we suggest that since $\rho_{\rm max}^{(\infty )}$ from Eq. (\ref{form}) is relatively easy to achieve and
has a universal form (independent on any qubit parameter or physical realization), it can also serve as a standard  of entanglement for quantum networks.
In other words, since entanglement is a measurable quantity, it should have a metrological standard associated: one can go further and speculate that the value of concurrence of 0.309 could then play the same role in metrology as other numbers do for the corresponding physical quantities ({\it e.g.} the temperature of the triple point of water, the value of the transition frequency between the two hyperfine states of Cs-133 atoms in atomic clocks, {\it etc.}).

\section{Conclusions}

In conclusion, we have shown that entanglement emerges in  a certain region of the parameter space of driven, coupled qubits  interacting with separate reservoirs; we give analytical results characterizing the stationary entanglement and we show how to further propagate this entanglement in a quantum network.

\section{Acknowledgments}
This work was supported by the Academy of Finland (Acad. Res. Fellowship 00857 and projects 129896 and 118122).

Note: After this work was completed we became aware of similar entanglement production effects in a number of systems under specific nonequilibrium conditions, for example in double quantum dots interacting {\it via} a bosonic environment and exposed to distinct fermionic baths \cite{doubleqdot},
in interacting spin gases under a random process which projects the particles on a specified state \cite{briegel},
in qubits with tunable Larmor frequencies and interaction strengths \cite{zhang},
in coupled electromagnetic fields and spin chains at stochastic resonance \cite{huelgaplenio}, and in coupled polaritons \cite{polaritons}.

\appendix
\section{Single qubit case}
\label{appendixa}

\begin{figure}[htb]
\includegraphics[width=9cm]{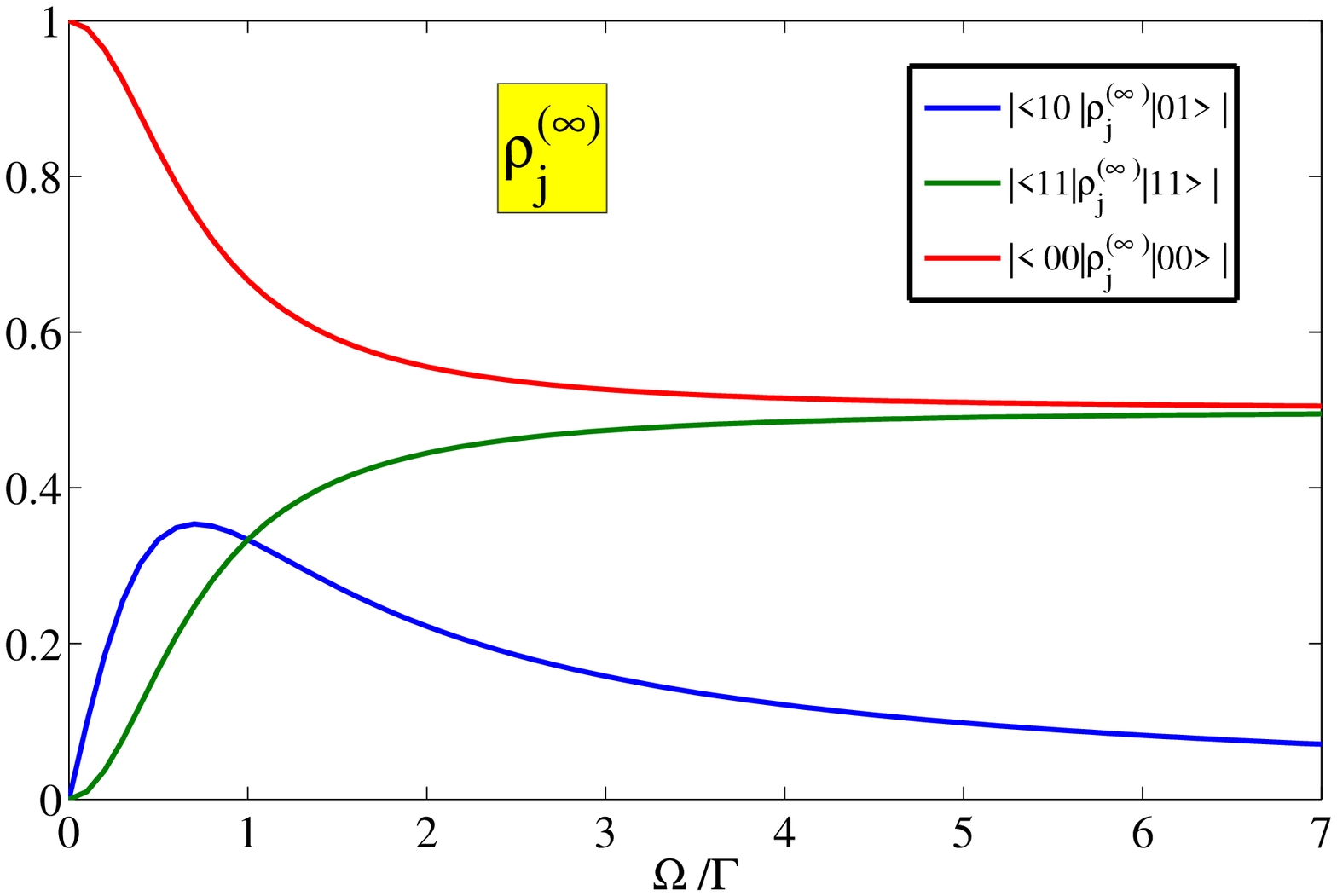}
\includegraphics[width=6cm]{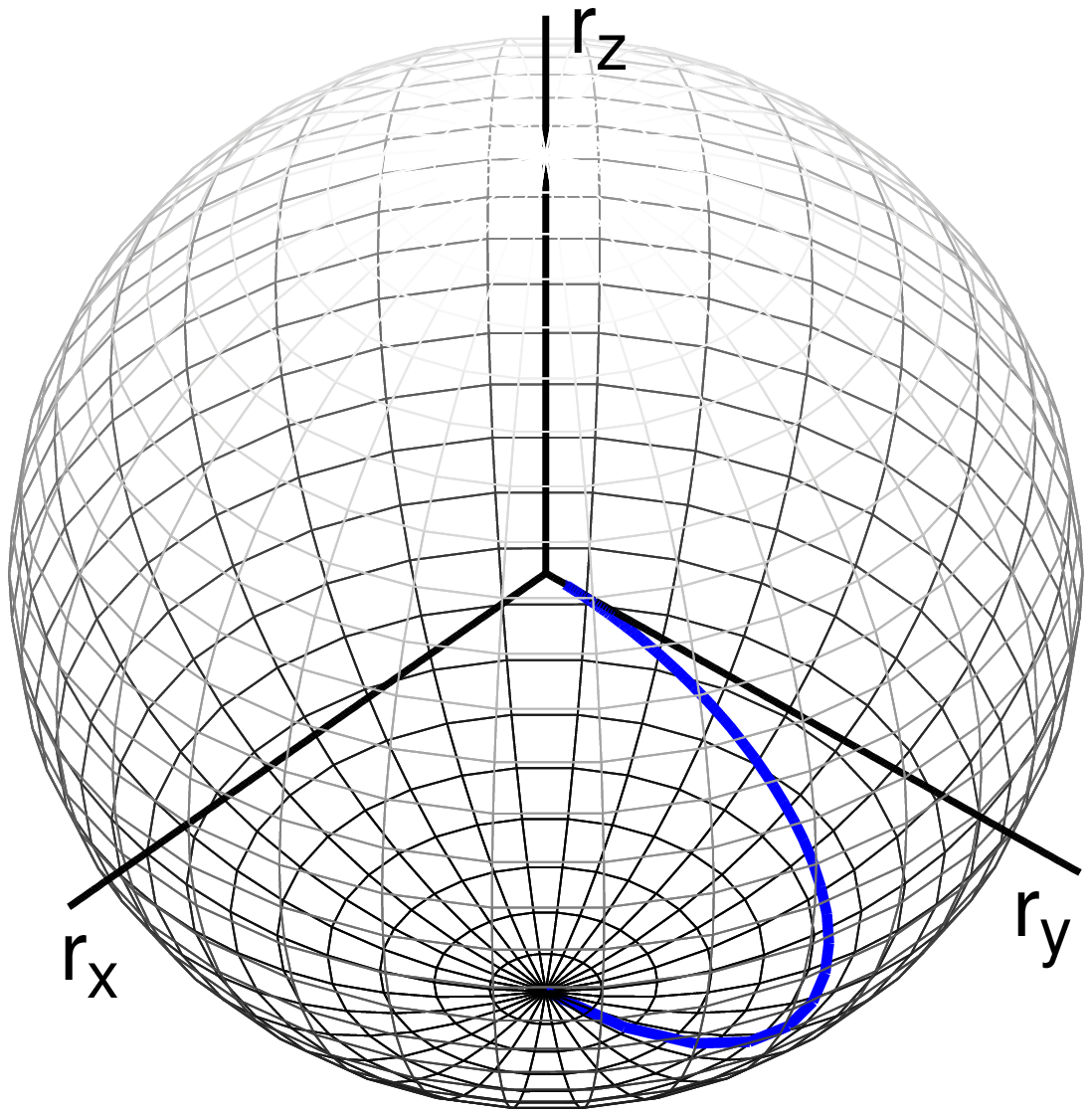}
\caption{The elements of the steady state of the single qubit as a function of $\Omega / \Gamma$ (left) and the corresponding representation (semicircle, in blue) on the Bloch sphere (right).}
\label{singlequbit}
\end{figure}

Here we review the case of one qubit under dissipation.  The rotating-frame single-qubit density matrix $\rho_{j}$ can be parametrized on the Bloch-sphere,
\begin{equation}
\rho_{j}= \frac{1}{2}(I+\vec{r}\vec{\sigma}),\label{eqna}
\end{equation}
where $\vec{r}$ is the Bloch vector of components $(r_{x},r_{y},r_{z})$, the norm of which indicates the purity of the state,
$Tr\rho_{rf}^2 = (1+|\vec{r}|^2)/2$.
The decoherence is described by Eq. (\ref{eqnnb}),
\begin{eqnarray}
\dot{\rho_{j}} = -\frac{i\Omega}{2}[\sigma^x,\rho_{j}]  + \frac{\Gamma}{2}\left( 2\sigma^-\rho_{j}\sigma^+ - \sigma^+\sigma^-\rho_{j} - \rho_{j}\sigma^+\sigma^- \right) .\label{eqnb}\ \
\end{eqnarray}
From Eq.(\ref{eqna},\ref{eqnb}) we obtain following kinetic equation for the components of the Bloch vector,
\begin{eqnarray}
\dot{r}_{z} = \Omega r_{y} - \Gamma(1+r_{z}) , \nonumber \ \
\dot{r}_{y} = -\frac{\Gamma}{2}r_{y} - \Omega r_{z}, \nonumber \ \
\dot{r}_{x} = -\frac{\Gamma}{2}r_{x}.\nonumber
\end{eqnarray}
The last equation has a simple solution, $r_{x}(t)=r_{x}(0)\exp (-\Gamma t/2)$.
The first two equations can be solved by introducing the notation $|\xi (t)\rangle = [r_{z},r_{y}]^{T}$ and
$|\chi \rangle = [-\Gamma , 0]$;
we find the equation
\begin{equation}
\frac{d}{dt}|\xi (t)\rangle = M|\xi (t)\rangle + |\chi\rangle .\label{equ}
\end{equation}
Here the matrix $M$ is defined by
\begin{equation}
M = \left(\begin{array}{cc} -\Gamma & \Omega \\ -\Omega & -\frac{\Gamma}{2} \end{array}\right),
\end{equation}
and has eigenvalues $\lambda_{1,2} = (1/4)(-3\Gamma \mp \sqrt{\Gamma^2 -16\Omega^2})$ corresponding to
(unnormalized) eigenvectors $|\xi_{1,2} \rangle = [(\Gamma \pm \sqrt{\Gamma^2 -16\Omega^2})/4\Omega , 1]^{T}$.
The stationary solutions of Eq. (\ref{equ}) are obtained for $t\gg \Gamma^{-1}$ as $|\xi ^{(\infty )}\rangle = -M^{-1}|\chi\rangle$,
resulting in the density matrix Eq. (\ref{hoi}),
\begin{eqnarray}
\rho_{j}^{ (\infty )} = \left(\begin{array}{cc} \frac{\Omega^2}{\Gamma^2 + 2\Omega^2 } & \frac{-i\Omega\Gamma}{\Gamma^2 + 2\Omega^2} \\
 \frac{i\Omega\Gamma}{\Gamma^2 + 2\Omega^2} & \frac{\Gamma^2 + \Omega^2}{\Gamma^2 + 2\Omega^2 }\end{array}\right).
\end{eqnarray}
The elements of this density matrix are shown in Fig. \ref{singlequbit}. In the Bloch sphere representation, the stationary state density matrix is parametrized by
$r_{x}^{(\infty )} = 0$, $r_{y}^{(\infty )} = 2\Omega\Gamma /(2\Omega^2 + \Gamma^2 )$,
and $r_{z}^{(\infty )} = - \Gamma^{2}/(\Gamma^{2} + 2\Omega^{2})$, describing a semicircle if radius $1/2$ in the $y-O-z$ plane, $(r_{z} + 1/2 )^{2} + r_{y}^{2}/2 = 1/4$ (shown in blue in the Bloch sphere of Fig. \ref{singlequbit}).

Depending on the values of $\Omega /\Gamma$ the eigenvalues $\lambda_{1,2}$ can have imaginary parts or can be real,
corresponding respectively
to underdamped and overdamped dynamics.
In the underdamped regime, defined by $\Omega /\Gamma >1/4$,
the elements of the density matrix oscillate (Rabi oscillations) before reaching the stationary state; deep in this regime
when  $\Omega /\Gamma \gg 1/4$, we find
\begin{eqnarray}
\rho_{j}^{(\infty )} = \left(\begin{array}{cc} 1/2 & -i\Gamma /2\Omega \\
 i\Gamma /2\Omega & 1/2 \end{array}\right)
\end{eqnarray}
In  the overdamped regime defined by $\Omega /\Gamma <1/4$ there are no Rabi oscillations; deep in this
regime, when $\Omega /\Gamma \ll 1/4$, we can write approximately
\begin{eqnarray}
\rho_{j}^{(\infty )} = \left(\begin{array}{cc} 0 & -i\Omega /\Gamma \\
 i\Omega /\Gamma & 1\end{array}\right).
\end{eqnarray}
Finally, the critically damped regime occurs at $\Omega = \Gamma /4$, and
\begin{eqnarray}
\rho_{j}^{(\infty )} = \left(\begin{array}{cc} 1/18 & -i 2/9 \\
 i 2/9 & 17/18 \end{array}\right).
\end{eqnarray}

From this, we see that for values of  $\omega^{xx}$ of the order of $\Gamma$ the emergence of stationary entangled states
 depends on having relatively large single-qubit coherences $r_{y}$, which happens not far from the
critically damped regime (the single-qubit coherence reaches a maximum of $1/2\sqrt{2}$ at $\Omega /\Gamma = 1/\sqrt{2}$).
 This is due to the fact that the interaction couples the qubits via the off-diagonal terms.

\end{document}